\documentclass[letterpaper,english,aps, prl, reprint, superscriptaddress]{revtex4-1}
\usepackage[T1]{fontenc}
\usepackage[latin9]{inputenc}
\usepackage{amsmath}
\usepackage{amssymb}
\usepackage{bbold}
\usepackage{graphicx}

\makeatletter

\pdfpageheight\paperheight
\pdfpagewidth\paperwidth

\PassOptionsToPackage{position=top}{subfig}
\usepackage{hyperref}
\hypersetup{
	colorlinks = true,
	allcolors = blue
}

\usepackage{times}

\makeatother

\usepackage{babel}
\begin{document}
\title{Exact new mobility edges between critical and localized states}

\author{Xin-Chi Zhou}
\thanks{These authors contribute equally to this work.}
\affiliation{International Center for Quantum Materials, School of Physics, Peking University, Beijing 100871, China}
\affiliation{Hefei National Laboratory, Hefei 230088, China}

\author{Yongjian Wang}
\thanks{These authors contribute equally to this work.}
\affiliation{School of Mathematics and Statistics, Nanjing University of Science and Technology, Nanjing 210094, China}
\affiliation{School of Mathematical Sciences, Laboratory of Mathematics and Complex Systems, MOE, Beijing Normal University, Beijing 100875, China}

\author{Ting-Fung Jeffrey Poon}
\thanks{These authors contribute equally to this work.}
\affiliation{International Center for Quantum Materials, School of Physics, Peking University, Beijing 100871, China}
\affiliation{Hefei National Laboratory, Hefei 230088, China}

\author{Qi Zhou}
\email{qizhou@nankai.edu.cn}
\affiliation{Chern Institute of Mathematics and LPMC, Nankai University, Tianjin 300071, China}

\author{Xiong-Jun Liu}
\email{xiongjunliu@pku.edu.cn}
\affiliation{International Center for Quantum Materials, School of Physics, Peking University, Beijing 100871, China}
\affiliation{Hefei National Laboratory, Hefei 230088, China}
\affiliation{International Quantum Academy, Shenzhen 518048, China}

\begin{abstract} 
The disorder systems host three types of fundamental quantum states, known as the extended, localized, and critical states, of which the critical states remain being much less explored. Here we propose a class of exactly solvable models which host a novel type of exact mobility edges (MEs) separating localized
states from robust critical states, and propose experimental realization. Here the robustness refers to the stability against
both single-particle perturbation and interactions in the few-body regime. The exactly solvable one-dimensional models are featured by quasiperiodic
mosaic type of both hopping terms and on-site potentials. The analytic results enable us to \textit{unambiguously} obtain the critical states which otherwise require arduous numerical verification including the careful finite size scalings. 
The critical states and new MEs are shown to be robust,
illustrating a generic mechanism unveiled here that the critical states are protected by zeros of quasiperiodic hopping terms in the thermodynamic limit. Further, we propose a novel experimental scheme to realize the exactly solvable model and the new MEs in an incommensurate Rydberg Raman superarray. This work may pave a way to precisely explore the critical states and new ME physics with experimental feasibility.
\end{abstract}

\maketitle
\textcolor{blue}{\em Introduction.}--Anderson localization (AL) is a fundamental
and ubiquitous quantum phenomenon that quantum states are exponentially localized 
due to disorder~\citep{anderson1}. 
Scaling theory shows that all noninteracting states are localized in one and two dimensions with arbitrarily small disorder strength~\citep{scaling,scaling2}, 
while in three dimension (3D), the localized
and extended states can coexist at finite
disorder strength, and be separated by a critical energy $E_{c}$, dubbed the mobility edge (ME). The
ME leads to various fundamental phenomena, such as metal-insulator
transition by varying the particle
number density or disorder strength~\citep{anderson2}. Moreover, a system 
with ME exhibits strong thermoelectric response, enabling application
in thermoelectric devices~\citep{thermoe1,thermoe2,thermoe3}. An important feature of ME between extended and localized states is that it is stable, and can survive under perturbations and interactions~\citep{interaction1,interaction2,interaction3,interaction4,cexp3-1}.

Unlike in randomly disordered system, the extended-AL transition and ME can exist in 1D system with quasiperiodic potential~\citep{qp1,qp2,qp2-1,qp3,qp4,qp5,qp6,qp7,qp8,qp8-1,qp9,qp9-1,qp11,qp11-1,qp11-2,qp11-3,qp12,qp14}. This result has triggered lots of experimental studies in realizing quasiperiodic systems with 
ultracold atoms~\citep{cexp1,cexp1-1,cexp1-2,cexp1-3,cexp1-4,cexp1-5,cexp1-6,cexp2,cexp3,cexp4,cexp5} and other systems like photonic
crystals, optical cavities, and superconducting circuits~\citep{exp1,exp2,exp3,exp4-1,exp4,exp5}. More importantly, quasiperiodic systems can host a third type of states called critical
states~\citep{exp5,critical1,EAA,critical2,critical3,critical4}. The critical states are extended but non-ergodic, locally scale-invariant and fundamentally
different from the localized and extended states in spectral statistics~\citep{spec1,spec2,spec3}, multifractal properties~\citep{mf1,mf2,mf3},
and dynamical evolution~\citep{diff1,diff2,diff3}. With interactions, the single-particle critical states may become many-body
critical (MBC) phase~\citep{qp12,mbc1,mbc2} that interpolates the
thermal and many-body localized phase~\citep{mbl1,mbl2}. However, unlike localized and extended states, to confirm critical states is more subtle and requires arduous numerical calculations like finite-size scaling. It remains unclear what generic mechanism leads to the critical states. Therefore, it is highly important to develop exactly solvable models with critical states being unambiguously determined and fully characterized. Moreover, similar to the ME for extended and localized states, are there new MEs separating critical from
localized states~\citep{qp11}, in particular, in experimentally feasible models? A definite answer to this fundamental question is yet elusive but may be provided by resolving the following issues. First, one can develop exactly solvable models with analytic MEs between critical and localized states. Further, one needs to confirm that such new MEs are robust, e.g. in the presence of perturbation and/or interactions. Finally, the proposed exactly solvable models are feasible in experimental realization.

In this Letter, we propose a class of exactly solvable 1D models featured with
mosaic type quasiperiodic hopping coefficients and on-site potential, and obtain \textit{unambiguously} critical states and robust exact
MEs. The new MEs fundamentally distinct from those in previous exact solvable models~\cite{qp9}. The localization and critical features of all quantum states in the spectra are precisely determined by extending Avila's global theory~\citep{avila}, enabling an accurate characterization of the critical states and new MEs. We further confirm the robustness of MEs against single-particle perturbation and interactions in the few-body regime. The robustness is rooted in a profound mechanism unveiled with our exactly solvable models that the critical states are protected by incommensurately distributed zeros of mosaic hopping terms in thermodynamic limit. Finally, we propose a novel scheme with experimental feasibility to realize and detect the exact MEs in {\em Raman superarray} of Rydberg atoms.

\begin{figure}[tp]
\includegraphics{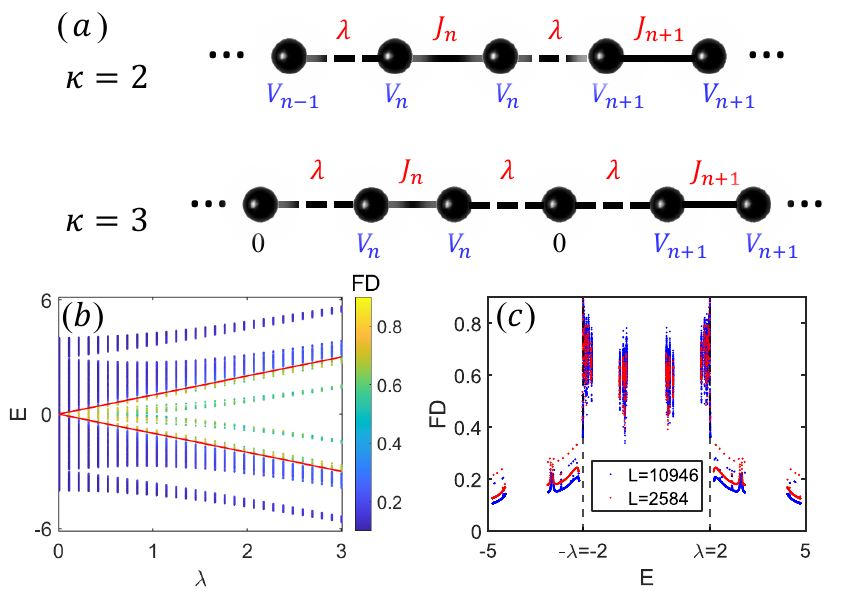}

\caption{\label{FD}(a) The 1D quasiperiodic mosaic model with $\kappa=2$ and $\kappa=3$. The black solid and dashed lines denote the quasiperiodic hopping ($t_j=J_n$) and constant hopping ($t_j=\lambda$), respectively. The black sphere denotes a lattice site, and $V_n$ is the quasiperiodic mosaic potential. Here $J_n=V_n=2t_0\cos(2\pi\kappa\alpha n)$, $n$ is an integer. (b) Fractal dimension (FD) of different eigenstates as a function of corresponding energies $E$ and $\lambda$ for $L=2584$. The red lines denote the MEs $E_c=\pm\lambda$. (c) FD versus $E$ with fixed $\lambda=2.0$ for $L=2584$ (red dots) and $L=10946$ (blue dots). The dashed lines represent the MEs. The $t_0$ is set to 1 as energy unit.}

\end{figure}

\textcolor{blue}{\em Model.}--We propose a class of quasiperiodic mosaic models as pictorially shown in Fig.~\ref{FD}(a), and described by
\begin{equation}
H=\sum_{j}(t_{j}a_{j}^{\dagger}a_{j+1}+\mathrm{h.c.})+\sum_{j}V_{j}n_{j},\label{eq:Ham}
\end{equation}
where the particle number operator $n_{j}=a_{j}^{\dagger}a_{j}$, with $a_{j}^{\dagger} (a_{j})$ the creation (annihilation) operator at site $j$, and
the key ingredients in the models are that both the quasiperiodic hopping coefficient $t_{j}$ and on-site potential $V_{j}$ are mosaic, with
\begin{equation}
\{t_{j}, V_{j}\}=\begin{cases}
\{\lambda,\ 2t_0\cos[2\pi\alpha(j-1)+\theta]\}, & j=1\ \mathrm{mod\,}\kappa,\\
2t_0\cos(2\pi\alpha j+\theta)\{1,\ 1\}, & j=0\ \mathrm{mod\,}\kappa,\\
\{\lambda,0\}, & \mathrm{else}.
\end{cases}\label{eq:mosaic}
\end{equation}
Here $\kappa$ is an integer and $\kappa\geq2$. $\lambda$
and $\theta$ denote hopping coefficient and phase offset, respectively. We take for convenience $t_0=1$, and set $\theta=0$ and $\alpha=\lim_{n\rightarrow\infty}(F_{n-1}/F_{n})=(\sqrt{5}-1)/2$ without affecting generality,
with $F_{n}$ being Fibonacci numbers. 
For finite system one may choose the system size $L=F_{n}$ and $\alpha=F_{n-1}/F_{n}$
to impose the periodic boundary condition for numerical diagonalization
of the tight-binding model in Eq. (\ref{eq:Ham}). To facilitate our discussion we focus on the minimal model for $\kappa=2$ in main text. The results with $\kappa>2$ are put in Supplemental Material~\cite{SM}. We shall prove that the minimal model has exact energy-dependent MEs separating
localized states and critical states, which are given by
\begin{equation}
E_c=\pm\lambda.
\end{equation}
Before showing the rigorous proof, we present numerical verification. For this exactly solvable model, the different types of states can be identified by the fractal dimension
(FD), defined for an arbitrary $m$-th eigenstate $|\psi_{m}\rangle=\sum_{j=1}^{L}u_{m,j}a_{j}^{\dagger}|vac\rangle$
as $\mathrm{FD}=-\lim_{L\rightarrow\infty}\ln(\mathrm{IPR})/\ln(L)$,
with the inverse participation ratio (IPR) being $\mathrm{IPR}=\sum_{j}|u_{m,j}|^{4}$.
The FD tends to 1 and 0 for the extended and localized states, respectively,
while $0<\mathrm{FD}<1$ for critical states. Fig.~\ref{FD}(b) shows $\mathrm{FD}$ as a function of $\lambda$
for different eigenstates of
eigenvalues $E$. The red lines starting from band center denote the MEs $E_c=\pm\lambda$, across which $\mathrm{FD}$ changes from values 
close to $0.5$ to values close to 0, indicating a critical-to-localization
transition predicted by the analytic results. Particularly, we fix
$\lambda=2.0$ and show $\mathrm{FD}$ of different eigenstates in Fig.~\ref{FD}(c)
as a function of the corresponding eigenvalues for different sizes. The dashed lines in the figure are the MEs $E_c=\pm\lambda=\pm2.0$.
One can observe that the fractal dimension $\mathrm{FD}$ tends to $0$ for
all states in energy zones with $|E|>\lambda$ with increasing
the system size, implying that those states are localized. On the
contrast, in energy zones with $|E|<\lambda$, the $\mathrm{FD}$ magnitude is far different from 0 and 1, and nearly independent of the system size. A more careful finite size scaling for FD can be found in~\cite{SM}.

\textcolor{blue}{\em Rigorous proof.}--The MEs of the models in Eqs.~\eqref{eq:Ham}-\eqref{eq:mosaic} can be analytically obtained by computing  Lyapunov exponent (LE) $\gamma_{\epsilon}$ in combination with zeros of hopping coefficients, which provides the unambiguous evidence of the critical zone. 
Denote $T_i$ to be the one-step transfer matrix of the Schr{\"o}dinger operator at site $i$, i.e. $(C_{i+1}, C_i)^\top = T_i (C_{i}, C_{i-1})^\top$ and $\mathcal{T}_{i} = T_i T_{i-1} \cdots T_1$.
The LE $\gamma_0$ for a state with energy $E$ is computed via $\gamma_{\epsilon}(E)=\lim_{m\to\infty}\int d\theta\ln\|\mathcal{T}_{m}(\theta+i\epsilon)\|/(2\pi m)$,
where $\|\cdot\|$ denotes the norm of matrix and $\epsilon$ is imaginary part of complexified $\theta$. We will extend Avila's global theory~\citep{avila} to singular cocycles, and  show that $\gamma_0=\lim_{\epsilon\to\infty} \gamma_\epsilon = \kappa^{-1}\ln \|\lim_{\epsilon\to\infty} \mathcal{T}_\kappa(\theta+i\epsilon)\|$~\citep{explainEq}. In particular for $\kappa=2$,
\begin{equation*}
    \mathcal{T}_2(\theta+i\epsilon)=\frac{1}{\lambda M}\begin{pmatrix}
E-M & -M \\
\lambda & 0
\end{pmatrix}\begin{pmatrix}
E-M & -\lambda \\
M & 0
\end{pmatrix},
\end{equation*}
where $M=2\cos(2\pi\alpha+\theta+i\epsilon)$ so that the LE is given by
\begin{equation}
    \gamma_0(E)= \frac{1}{2}\ln\left||E/\lambda|+\sqrt{(E/\lambda)^{2}-1}\right|.
\end{equation}
For $|E|>|\lambda|$, one has $\gamma_{0}(E)>0$ and the state associated with $E$ is localized with the localization length $\xi(E) = \gamma_0^{-1}$.
If $|E|<|\lambda|$, $\gamma_{0}(E)=0$ and for such LE the eigenstates can in general be either extended or critical, which belong to absolutely continuous (AC) spectrum or singular continuous (SC) spectrum, respectively~\cite{avila2017}. There are two basic approaches to rule out the existence of AC spectrum (extended states), one is introducing unbounded spectrum~\cite{qp11} and the other is introducing zeros of hopping terms in Hamiltonian~\cite{simon,J}. For our model, there exists a sequence of sites
$\{2j_{k}\}$ such that $t_{2j_{k}}\rightarrow0$ in the thermodynamics limit, so there is no AC spectrum (extended states) \cite{noACreason}, and the eigenstates associated with $|E|\leq|\lambda|$ are all critical. In summary, vanishing LEs $\gamma_0=0$ and zeros of incommensurate hopping coefficients altogether unambiguously determine the critical region for $|E|\leq|\lambda|$ and positive LEs determine the localized region for $|E|>|\lambda|$. Therefore $E=\text{\ensuremath{\pm\lambda}}$ mark critical energies separating localized states and critical states, manifesting MEs.

\begin{figure}[tp]
\includegraphics{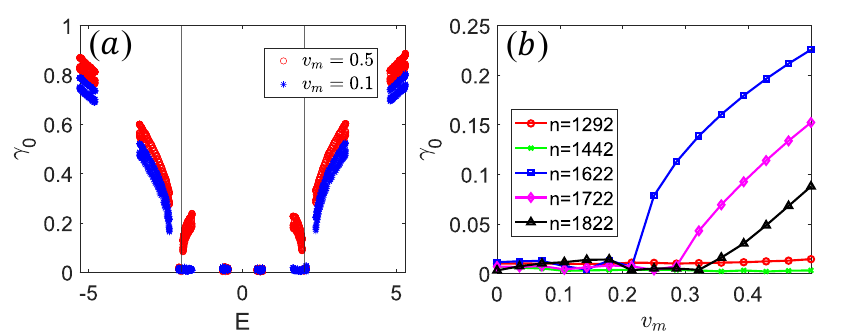}\caption{\label{fig:perturbation}Numerical LEs $\gamma_0$ for perturbation strength $v_m$. (a) $\gamma_0$ as a function of eigenvalues $E$ for different $v_m$. Solid lines are original MEs. (b) $\gamma_0$ as a function of $v_m$ for eigenstates $\lvert \psi_{n}\rangle$ from the band center ($n=1292$) to the states nearby original MEs ($n=1822$). Critical states nearby MEs can be driven to localized states while critical zones nearby band center remains unchanged. The other parameters are $\lambda=2.0$ and $L=2584$.}
\end{figure}

\textcolor{blue}{\em Mechanism of critical states.}--The emergence of MEs and critical states has a universal underlying mechanism unveiled from the exact results. 
Namely, it is due to incommensurately distributed zeros of hopping coefficients~\cite{incommenZeros} in the thermodynamic limit and vanishing the LE. 
Such zeros in the hopping coefficient $t_j$ 
effectively divide system into weakly coupled subchains, ruling out possibility of supporting extended states, and leaving the localized or critical states depending on the corresponding LEs. To further verify there is a plethora of critical states in the whole spectrum, let us consider first the special case with vanishing hopping coefficient $\lambda\rightarrow0$ [Fig.~\ref{FD}(a)]. In this case the model is divided into a series of dimmers, and each dimmer renders a $2\times2$ matrix with all elements being $J_j=V_j=2\cos(2\pi\alpha j)$. Then the eigenvalues are simply $E_1=2J_j$ and $E_2=0$, of which the former corresponds to the 
localized states, 
and the latter represents a zero-energy flat band whose degeneracy equals to half of system size. Note that a linear combination of the zero modes are also eigenstates of the model, which can be either localized, extended or critical.
Further inclusion of $\lambda$ hybridizes the zero-energy flat-band modes and localized modes, yielding the critical states and MEs between them and localized ones. This mechanism also explains emergence of critical states starting from band center. Moreover, the number of critical states equals to that of localized states under the exactly solvable condition, 
as verified by the numerical counting.

This zero hopping coefficient mechanism also explains a novel feature that 
the critical zone is robust against single-particle perturbation which tunes the model away from exactly solvable condition. We consider an extra mosaic on-site potential term $V_p$ as perturbation, which represents the mismatch between mosaic hopping and onsite potentials in Hamiltonian given in Eq.~\eqref{eq:mosaic}, and is relevant to real experiment,
\begin{equation}
H_{p}=H+\sum_{j}V^{p}_{j}n_j,
\end{equation}
where $V^{p}_{j}=v_{m}\cos(2\pi\alpha j+\theta)$ for even-$j$ sites and $V^{p}_{j}=v_{m}\cos[2\pi\alpha(j-1)+\theta]$) for odd-$j$ sites. Fig.~\ref{fig:perturbation} shows numerically calculated LEs for different $v_m$, with $\lambda=2$. The sufficiently strong mismatch potentials can drive critical states nearby MEs into localized ones while for critical zones nearby the band center the states remains unchanged [Fig.~\ref{fig:perturbation}(a)]. A more careful investigation of LEs 
shows that to drive critical states into localized ones requires finite $v_m=v_m^c$, 
with the magnitude $v_m^c$ depending on the location of the critical state [Fig.~\ref{fig:perturbation}(b)]. These results manifest the robustness of critical zones against perturbations, which is a key point of the present model. 
In comparison, the celebrated Aubry-Andr\'e (AA) model exhibit a self-duality point at $V=2t$, at which all states are critical~\cite{qp1}. However, those critical states are not robust and shall be quenched to localized state for infinitesimal perturbations. The present study unveils a generic mechanism to obtain robust critical zones protected by the zeros of hopping coefficients with vanishing LEs in the thermodynamic limit, which are not removed by the perturbations. This also shows a nontrivial regime that while Avila's global theory cannot give analytical MEs, the MEs exist.

\begin{figure}[tp]
\includegraphics{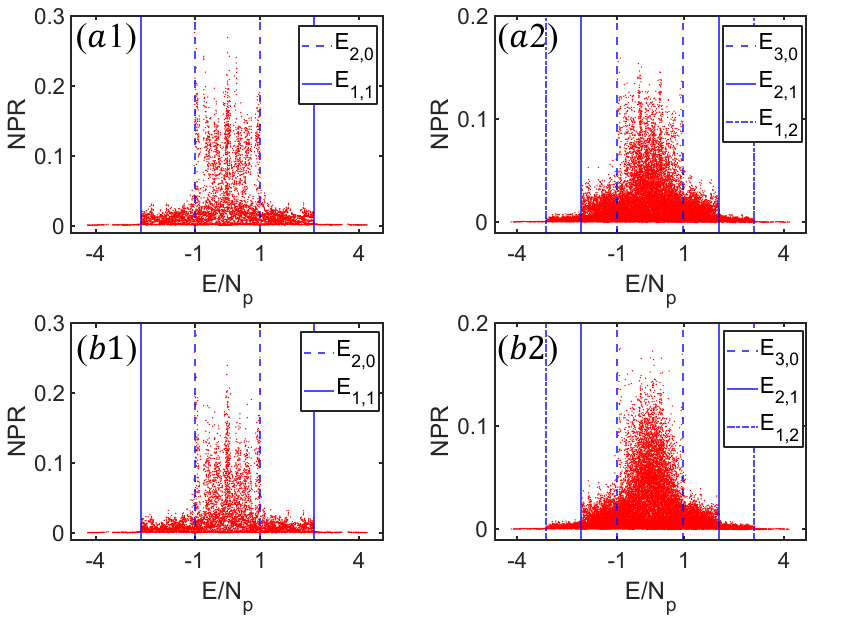}\caption{\label{fig:NPR}Normalized participation ratio (NPR) as a function
of energy density $E/N_{\mathrm{p}}$ of few hardcore bosons with $U=0$ (upper panel) and $U=1.4$ (lower panel), with $\lambda=1.0$ and $t_0=1.0$. Critical energy \textcolor{blue}{$E_{N_{c},N_l}$} is the maximal allowed energy for \textcolor{blue}{$N_{c}$} particles filled in critical orbitals and \textcolor{blue}{$N_{l}$} particles filled in localized orbitals.
(a1)(b1) $N_{\mathrm{p}}=2$, $L=120$. (a2)(b2) $N_{\mathrm{p}}=3$, $L=60$. The critical energy $E_{2,0}$($E_{3,0}$) is equal to the single particle ME $E=\lambda$ for two(three)-boson case.}
\end{figure}

\textcolor{blue}{\em Robustness against interactions.}--We further demonstrate the robustness of MEs in the presence of interactions by studying a few-body Hamiltonian 
given by
\begin{equation}
H=H_0+U\sum_j n_j n_{j+1},
\end{equation}
where $H_0$ denotes the system with few hard-core bosons with Hamiltonian in Eq.~\eqref{eq:Ham}, with $\langle n_j\rangle\leq1$, and $U$ denotes the strength of neighboring interactions. 
This model can be simulated with Rydberg atoms (see details in next section).
We propose normalized participation ratio (NPR) to detect the MEs in the few-body system. The NPR of an eigenstate $\left|\psi_m\right>=\sum_c u_{m,c}\left|c\right>$ is defined as $\text{NPR} = 1/\left(V_H \sum_c\left|u_{m,c}\right|^4\right)$, where $\{c\}$ is the computational basis and $V_H$ is the size of the Hilbert state~\cite{mbl3}.
When $U=0$, the few-body states are product states of single particle orbitals. In the presence of single particle MEs (SPMEs), the few-body states can be categorized into three types~\citep{interaction1,interaction2}: all the particles occupy localized (critical) orbitals, and mixed
states with some particles in localized orbitals and others in critical orbitals. Denote the maximum energy of critical (localized) orbitals as $\lambda$ ($E_\mathrm{max}$), where $E_\mathrm{max}$ is the maximum energy of spectrum, we can construct the maximally allowed energy for $N_c$ ($N_l$) particles filled in critical (localized) orbitals as $E_{N_{c},N_{l}}=\left(N_{c}\lambda+N_{l} E_{\mathrm{max}}\right)/N_p$ with $N_p=N_c+N_l$. Then $E_{N_{c},N_{l}}$ locates the transition of different types of few-body states, where NPR changes discontinuously.
For instance, when $N_p=3$, the maximal allowed energy for a mixed state with 2 particles filled in critical orbitals and 1 particle filled in localized orbitals is $E_{2,1}$.
As shown in Fig.~\ref{fig:NPR}(a), NPR displays clear discontinuities at $E_{N_{c},N_{l}}$ as expected.

Our key observation is that the sharp discontinuities near $\pm E_{N_{{c}},N_{l}}$ persist for $U\neq0$ for the few-body regime, manifesting the robustness of SPMEs against few-body interactions. As shown in Fig.~\ref{fig:NPR}(b), for $U=1.4$, $E_{N_{{c}},N_{l}}$ can still identify the NPR transition. This is because only those states with at least two particles occupying neighboring sites can be
 influenced by the interaction, and the portion of such states is of order $\mathcal{O}(L^{-1})$, with $L$ the system size. Thus for relatively large $L$ almost all eigenstates are still product states of single-particle orbitals, except for the small portion affected by interaction. Note that the localized orbitals have zero contribution to NPR in large $L$ limit, and the number of critical orbitals determine NPR of the few-body states. Thus the NPR exhibits sharp transitions across critical energies of $E_{N_{c},N_{l}}$, which are related to single particle MEs,
showing the robustness of MEs and critical zones in the few-body case.
This novel result motivate us to realize the exactly solvable model and observe our predictions with Rydberg atoms arrays~\cite{tweezer1,tweezer2,rydberg1} which are natural platforms to simulate hard-core bosons~\cite{hardcore1,hardcore2}.

\begin{figure}[tp]
\includegraphics{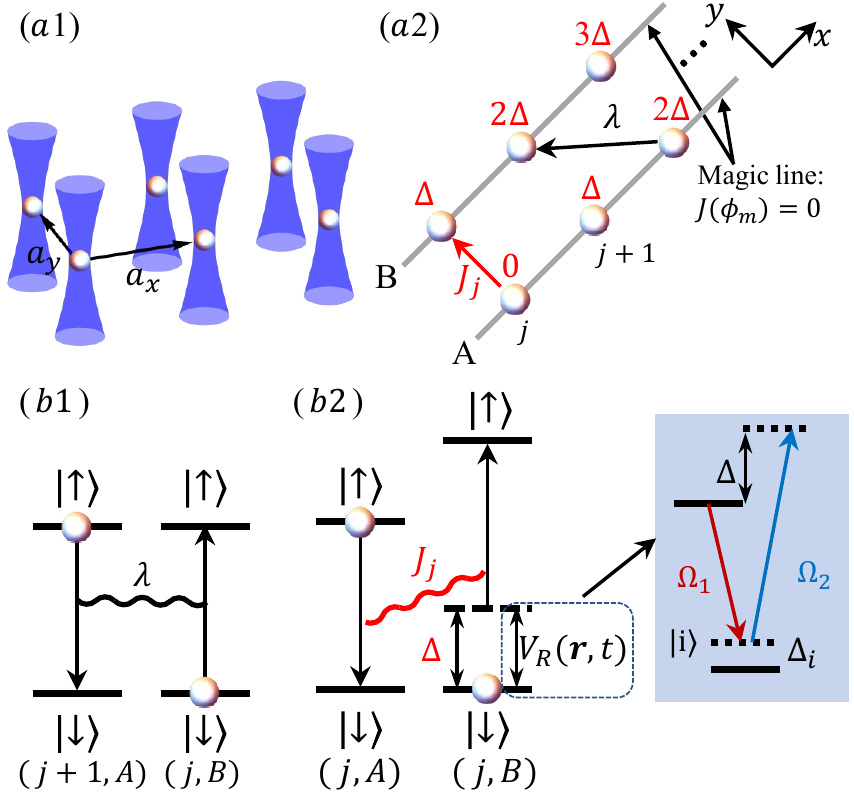}\caption{\label{fig:exp}Realization of the quasiperiodic mosaic model with $\kappa=2$ in incommensurate Raman superarrays of Rydberg atoms. (a1) Rydberg atoms trapped in optical tweezers to form a two-leg array. (a2) The equivalent two-leg lattice model. The incommensurate and constant hopping coefficients are denoted as $J_j$ and $\lambda$. An effective Zeeman splitting  gradient (red) modifies energy difference between Rydberg states. The Rydberg atoms are tapped along the magic angle $\phi_m$, at which the amplitude of dipolar interaction vanishes. (b1) The intrinsic dipole-dipole interaction can induce constant hopping coupling $\lambda$. (b2) The inter-leg exchange coupling between $A_j$ and $B_j$ is suppressed by the Zeeman detuning. A two-photon Raman process (see inset) compensates this energy penalty and induces the laser-assisted dipole-dipole interaction. The compensated exchange coupling realizes incommensurate hopping coupling $J_j$.
}
\end{figure}

\textcolor{blue}{\em Experimental realization}.--Finally we
propose an experimental scheme dubbed {\em incommensurate Raman superarray} of Rydberg atoms [Fig.~\ref{fig:exp}(a1,a2)] to realize the model in Eq.~\eqref{eq:Ham} with $\kappa=2$. We see that the realization of the Hamiltonian is precisely mapped to
the realization of a two-leg lattice model, with even (odd) sites mapped to the sites on $A$($B$)-leg as pictorially shown in Fig.~\ref{fig:exp}(a2), whose Hamiltonian reads
$H=\sum_{j}(J_{j}a^{\dagger}_{j}b_{j}+\lambda a^{\dagger}_{j}b_{j+1}+\mathrm{H.c.})+\sum_{j}V_{j}(a^{\dagger}_{j}a_{j}+b^{\dagger}_{j}b_{j})$. This basic idea can be directly generalized to realize models with larger $\kappa$ by introducing more legs in supperarray. The mapped Hamiltonian can be realized by Rydberg atoms based on three key ingredients. (i) The AB-leg superarray has an effective Zeeman splitting gradient applied in the $x$ direction (a2); (ii) two types of nearest neighbour couplings, with constant hopping coupling $\lambda$ simulated by intrinsic dipole-exchange interaction and quasi-periodic hopping coupling $J_j$ induced by laser-assisted dipolar interaction~\cite{laddi}; and (iii) an on-site incommensurate chemical potential $V_j$. Two Rydberg states $\lvert \downarrow\rangle =\lvert 70,S\rangle\equiv\lvert 0\rangle$ and $\lvert \uparrow\rangle =\lvert 70,P\rangle\equiv\lvert 1\rangle$ are chosen to simulate empty and occupied states at each site. 
As illustrated in Fig.~\ref{fig:exp}(b1), the intrinsic dipole-dipole interactions between two Rydberg states lead to an exchange coupling, which maps to constant hopping $\lambda$ of the hard-core bosons~\cite{hardcore1,hardcore2}.

The AB-leg superarray and laser-assisted dipole-dipole interactions altogether realize the incommensurate hopping $J_j$. The inter-leg exchange couplings between $A_j$ and $B_j$ sites are suppressed by a large energy detuning $\Delta$,
but can be further restored by applying the Raman coupling potential $V_{R}\propto\cos(4\pi\alpha j_{x})\cos(\pi j_{y})e^{i(\omega_{2}-\omega_{1})t}\sigma_{j_{x},j_{y}}^{x}+\mathrm{H.c.}$,
which is generated by two Raman beams $\Omega_{1,2}$ with frequency difference $\omega_{1}-\omega_{2}\approx\Delta$
such that the Zeeman splitting can be compensated by two-photon process [Fig.~\ref{fig:exp}(b2)]. Further, the spatial modulation of the Raman potential determines the incommensurate strength of the induced exchange couplings as
$J_{j}=2t_0\cos4\pi\alpha j$ (see Supplemental Material for details~\cite{SM}). To prohibit laser-assisted exchange couplings along $x$ direction, we use the angular dependence of dipole-dipole interaction $V_{\mathrm{dd}}=d^{2}(1-3\cos^{2}\phi)/R^{3}$
with $d$ and $R$ being the dipole moment and distance between two
Rydberg levels, respectively. There exists a "magic angle" $\phi_{m}=\arccos(1/\sqrt{3})\approx54.7^{\circ}$~\cite{angle,rydberg1}, along which the exchange coupling vanishes. By arranging the Rydberg atoms along the magic angle, we manage to prohibit coupling in the $x$ direction. Finally, the incommensurate mosaic potential $V_j$ can be realized via AC Stark shift~\cite{SM}. Adding up those ingredients together 
we reach the target model. The non-interacting critical states and MEs can be observed from the spectrum when a single hard-core boson is excited in this scheme, while the critical energies depicted in Fig.~\ref{fig:NPR} will be observed when several hard-core bosons are excited in experiment.

\textcolor{blue}{\em Conclusion and discussion.}--We have proposed a class of exactly solvable
1D incommensurate mosaic models 
to realize new and robust MEs separating critical states from localized states, and further proposed a novel experimental realization through incommensurate Raman superarrays of Rydberg atoms. 
The robust critical states and MEs originate from a combination of zeros of quasiperiodic hopping coefficients in the thermodynamic limit and the zero Lyapunov exponents (LEs), which can be analytically obtained for the proposed models and in agreement with the numerical studies. We note that these two features, serving as a generic mechanism, can provide the guidance to construct broad class of analytic models hosting robust critical states. Moreover, we demonstrate the robustness of MEs and propose NPR as a new probe to detect the MEs in the few-body regime. A future intriguing issue is to explore the interacting effects in the finite filling regime, which might lead to exotic many-body new MEs. Our work broadens the concept of MEs and provides a feasible lattice model that hosts exact MEs and unambiguous critical zones with experimental feasibility.

We thank Yucheng Wang for fruitful discussions. This work was supported by National Key Research and Development Program of China (2021YFA1400900 and 2020YFA0713300), the National Natural Science Foundation of China (Grants No. 11825401, No. 12261160368, No. 12071232, and No. 12061031), and the
Innovation Program for Quantum Science and Technology (Grant No. 2021ZD0302000). Q. Zhou was also supported by the Science Fund for Distinguished Young Scholars of Tianjin (No. 19JCJQJC61300) and Nankai Zhide Foundation.

\renewcommand{\thesection}{S-\arabic{section}}
\setcounter{section}{0}  
\renewcommand{\theequation}{S\arabic{equation}}
\setcounter{equation}{0}  
\renewcommand{\thefigure}{S\arabic{figure}}
\setcounter{figure}{0}  
\renewcommand{\thetable}{S\Roman{table}}
\setcounter{table}{0}  
\onecolumngrid \flushbottom 

\newpage

\begin{center}\large \textbf{Supplementary Material} \end{center}
This Supplemental Material provides additional information for the main text. In Sec.~\ref{LyapunovExponent}, we give the details of computing Lyapunov exponent. The numerical results for larger $\kappa$ are shown in Sec.~\ref{kappa}. In Sec.~\ref{FiniteSize}, we study finite size scaling of mean fractal dimension of our model. In Sec.~\ref{MorePerturbation}, we present other forms of perturbation potentials and other values of $\lambda$ in few-body calculations. In Sec.~\ref{LocLen}, we give numerical results of localization length. Finally, we give details of experimental realization in Sec.~\ref{ExpReal}.

\section{Lyapunov exponent}\label{LyapunovExponent}
In this section, we give a detailed derivation of Lyapunov exponent (LE) of the proposed model for arbitrary $\kappa$. We let $t_0=1$ for simplicity. For an irrational $\alpha$, consider the quasiperiodic Schr\"{o}dinger operators for the proposed model,
\begin{equation}\label{eq1} 	
(H_{\lambda,\alpha,\theta}\psi)_n\equiv V_n\psi_n+t_n\psi_{n+1}+t^*_{n-1}\psi_{n-1},
\end{equation}
with 
\begin{equation}
V_n(\theta)=\begin{cases}
2\cos[2\pi(n-1)\alpha+\theta]
   & \text{for } n=1 \text{ mod } \kappa,\\
2\cos(2\pi n\alpha+\theta)
   & \text{for } n=0 \text{ mod } \kappa,\\
0 & \text{otherwise,}
\end{cases}
\end{equation} 
\begin{equation}
t_n(\theta) = \begin{cases}
2\cos(2\pi n\alpha+\theta)
   & \text{for } n=0 \text{ mod } \kappa,\\
\lambda
   & \text{otherwise, }
\end{cases}
\end{equation}
$\theta \in \left[0,2\pi\right)$ and $\lambda \neq 0$. Denote the $m$-step transfer matrix of the operator as $\mathcal{T}_{m_0+m,m_0}(\theta) \equiv T_{m_0+m-1} T_{m_0+m-2} \cdots T_{m_0+1} T_{m_0}$, where $T_i$ is the one-step transfer matrix at site $i$ satisfying
\begin{equation}
\begin{pmatrix}
\psi_{i+1} \\\psi_i
\end{pmatrix} = T_i \begin{pmatrix}
\psi_i \\\psi_{i-1}
\end{pmatrix}
\end{equation}
and $T_i(\theta) = T_{i-\kappa}(\theta+2\pi\kappa\alpha)$. The LE $\gamma_0$ of our proposed model can be computed through the complexified LE
\begin{equation}
    \gamma_\epsilon(E) = \lim_{m\to\infty}\frac{1}{2\pi m}\int\ln\|\mathcal{T}_{m,1}(\theta+i\epsilon)\|d\theta,
\end{equation}
where $\|A\|$ denotes the norm of the matrix $A$, i.e. the square root of the largest eigenvalue of $A^\dagger A$.

Observe that the $\kappa$-step transfer matrix can be written as
\begin{equation*}
\begin{split} 	\mathcal{T}_{\kappa,1}(\theta+i\epsilon)&=\mathcal{T}_{2\kappa,\kappa+1}(\theta+i\epsilon-2\pi\alpha\kappa)=\mathcal{T}_{3\kappa,2\kappa+1}(\theta+i\epsilon-4\pi\alpha\kappa)=\cdots\\&=\frac{\begin{pmatrix}E/\lambda&-1\\1&0\end{pmatrix}^{\kappa-2}}{\lambda M(\theta+i\epsilon)}\begin{pmatrix} 		E-M(\theta+i\epsilon)&-M(\theta+i\epsilon)\\\lambda&0 	\end{pmatrix}\begin{pmatrix} 		E-M(\theta+i\epsilon)&-\lambda\\M(\theta+i\epsilon)&0 	\end{pmatrix}\\
&=\frac{1}{\lambda M}\begin{pmatrix}a_\kappa&-a_{\kappa-1}\\a_{\kappa-1}&-a_{\kappa-2}\end{pmatrix}\begin{pmatrix} 		E^2-2EM&-\lambda E+\lambda M\\\lambda E-\lambda M&-\lambda^2 	\end{pmatrix},
\end{split}
\end{equation*}
where $a_{\kappa}=\Delta^{-1}\left[((\beta+\Delta)/2)^{\kappa-1}-((\beta-\Delta)/2)^{\kappa-1}\right]$, $\Delta = \sqrt{\beta^2-4}$, $\beta=E/\lambda$ and $M(\theta+i\epsilon) = 2 \cos (\theta+i\epsilon)$.  The basic point here is that $\mathcal{T}_{\kappa,1}$ is  singular not analytic. While Avila's global theory is developed to one-frequency analytic cocycles \cite{A1}, here we will show how to extend it to deal with the singular setting.   

We observe that $\mathcal{T}_{\kappa,1}\in SL(2,\mathbb{R})$, and it is singular only at $\epsilon=0$, so that by Avila's global theory $\kappa \gamma_\epsilon$ is a convex, piecewise linear function with integer slopes for positive and negative $\epsilon$ separately, the main issue is the singular point $\epsilon=0$.
The key observation  is to consider 
$$\widetilde{T}_\kappa(\theta)\equiv M(\theta)\mathcal{T}_{\kappa,1}(\theta)=\frac{1}{\lambda }\begin{pmatrix}a_\kappa&-a_{\kappa-1}\\a_{\kappa-1}&-a_{\kappa-2}\end{pmatrix}\begin{pmatrix} 		E^2-2EM&-\lambda E+\lambda M\\\lambda E-\lambda M&-\lambda^2 	\end{pmatrix},$$ which are analytic about $\theta$. Thus, the Lyapunov exponent of the new matrix is defined by
$$\widetilde{\gamma}_\epsilon(E):=\lim_{m\to\infty}\frac{1}{2\pi m}\int\ln\|\widetilde{\mathcal{T}}_{m}(\theta+i\epsilon)\|d\theta,$$
Meanwhile, by definition of Lyapunov exponent 
\begin{equation}\label{lya}
	\kappa \gamma_\epsilon(E)= \widetilde{\gamma}_\epsilon(E)-\frac{1}{2\pi}\int_0^{2\pi}\ln| 2\cos (\theta+i\epsilon)|d\theta\end{equation}
where by classical Jenson's formula, one has 
	\begin{equation}\label{lya1}
\int_0^{2\pi}\ln| 2\cos (\theta+i\epsilon)|d\theta = 2\pi  |\epsilon|.
	\end{equation}
Since $ \widetilde{T}_\kappa(\theta)$ is analytic,  Avila's global theory \cite{A1} ensures $\widetilde{\gamma}_\epsilon(E)$ is a continuous, convex, piecewise linear function with respect to $\epsilon$. By \eqref{lya} and \eqref{lya1}, one thus conclude $\kappa \gamma_\epsilon$ is a convex, continuous, piecewise linear function with integer slopes for any $\epsilon\in \mathbb{R}$.

In the following, we will show that $\gamma_0(E)=\gamma_\epsilon(E)=\gamma_\infty(E)$. Let us complexify the phase of $\mathcal{T}_{\kappa}$, and let $\epsilon$ go to $+\infty$,
\begin{equation*}
	\mathcal{T}_{\kappa}(\theta+i\epsilon)=\begin{pmatrix}a_\kappa&-a_{\kappa-1}\\a_{\kappa-1}&-a_{\kappa-2}\end{pmatrix}\begin{pmatrix}-2\beta&1\\-1&0\end{pmatrix}+\mathcal{O}\left(e^{-\epsilon}\right)=\mathcal{T}_\kappa'+\mathcal{O}\left(e^{-\epsilon}\right)
\end{equation*}
so that
\begin{equation*}
	\begin{split} \kappa \gamma_\epsilon(E)
		=\ln\|\mathcal{T}'_\kappa\|+\mathcal{O}\left(\epsilon^{-1}\right).
	\end{split}
\end{equation*}
As we have proved, as a function of $\epsilon>0$, $\kappa\gamma_\epsilon(E)$ is a  convex, continuous, piecewise linear with integer slopes, which implies that, for all $\epsilon\in\mathbb{R}$
\begin{eqnarray}
	\gamma_\epsilon\equiv\frac{1}{\kappa}\ln\|\mathcal{T}_\kappa'\|=\frac{1}{\kappa}\ln\left||a_{\kappa+1}|+\sqrt{a_{\kappa+1}^2-1}\right|. \nonumber
\end{eqnarray}
In particular, $a_{1,2,3} = 0,1, \beta$ respectively so that for $\kappa=2$, the Lyapunov exponent 
$$\gamma_0 = 1/2\times\ln\left|\left(\left|E\right|+\sqrt{E^2-\lambda^2}\right)/\lambda\right|$$ as discussed in the maintext.

\begin{figure}[!h]
\includegraphics{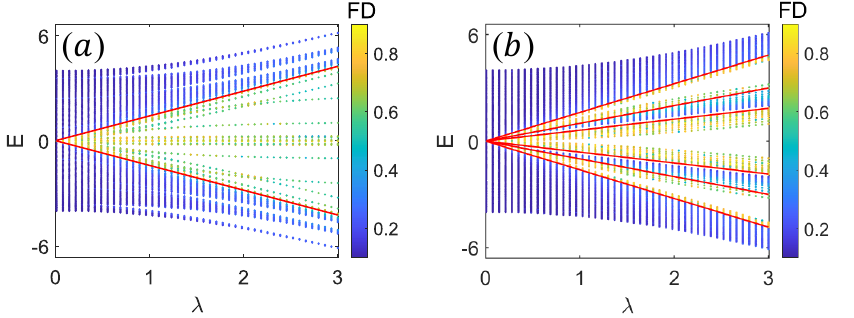}\caption{\label{fig:FDkappa}Fractal dimension ($\mathrm{FD}$) of corresponding eigenstates as a function of energies $E$ and constant hopping strength $\lambda$ for (a) $\kappa=3$ with lattice size $L=987$ and (b) $\kappa=4$ with lattice size $L=2584$. The red lines denote the MEs given by analytic results, and $t_0$ is set to 1 as energy unit.}
\end{figure}

\section{larger $\kappa$ case}\label{kappa}
In the main text, we have shown the calculated mobility edges (MEs) for the case $\kappa=2$. In this section, we present the analytic expression for the MEs with numerical verification for larger $\kappa$. The mobility edges occur at $ a_{\kappa+1}^2 = 1$, which simplifies to $\left(\csc\phi\sin\kappa\phi\right)^2 =1$, where $2\cos\phi = \beta$. Therefore the MEs are at
\begin{equation}
    E_{c}^{(\kappa)} = \pm 2\lambda\cos \frac{n_1\pi}{\kappa-1} \text{ or } \pm 2\lambda\cos \frac{n_2\pi}{\kappa+1},
\end{equation}
where $n_1 = 1, 2, \cdots, \lfloor(\kappa-2)/2\rfloor$ and $n_2 = 1, 2, \cdots, \lfloor\kappa/2\rfloor$. In particular for the cases of $\kappa=3$ or $4$, the MEs are given by
\begin{eqnarray}
    E_{c}^{(\kappa=3)}&=&\pm\sqrt{2}\lambda, \nonumber\\
    E_{c}^{(\kappa=4)}&=&\pm\lambda, \pm\frac{1+\sqrt{5}}{2}\lambda \text{ or } \pm\frac{1-\sqrt{5}}{2}\lambda. \nonumber
\end{eqnarray} \color{black}
As mentioned in main text, the $\mathrm{FD}$ tends to 1 and 0 for the extended and localized states, respectively, and $0<\mathrm{FD}<1$ for critical states. Fig.\ref{fig:FDkappa} shows the fractal dimension ($\mathrm{FD}$) as a function of $\lambda$ for different eigenstates of energies $E$. The red lines represent the MEs for $\kappa=3$ and $\kappa=4$, respectively. The $\mathrm{FD}$ changes from $0$ to value close to $0.5$ when the energy and mark the localization-to-critical transition predicted by analytic results. Both two cases agree with the analytic expressions as expected.

\section{Finite Size Scalings}\label{FiniteSize}
\begin{figure}[!h]
\includegraphics{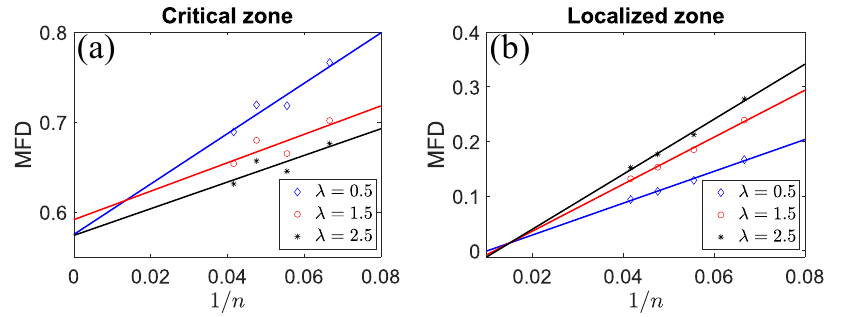}\caption{\label{fig:finitesize}$\mathrm{MFD}$ as a function of $1/n$ for the (a) critical zone and (b) localized zone with different $\lambda$. Here $n$ is the index of Fibonacci numbers $F_n$.}
\end{figure}

In this section, we study the fintie size scaling of mean fractal dimension (MFD) of critical zone and localized zone. We choose $\kappa=2$ case as an example. MFD of critical zone is the averaged fractal dimension of all eigenstates in critical regime, i.e., eigenstates with energies $|E|<|\lambda|$. MFD of localized zone can be obtained in a similar way. Fig.~\ref{fig:finitesize} (a) and (b) show the $\mathrm{MFD}$ as a function of $1/n$ for the critical zone and localized zone, respectively. It is observed that $\mathrm{MFD}$ approaches to finite value between 0 and 1 for the critical zone, and approaches to 0 for the localized zone.

\begin{figure}[!h]
\includegraphics{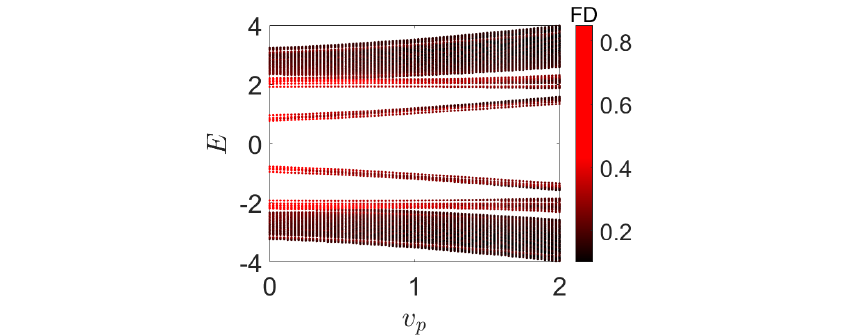}\caption{\label{fig:FDSMonsite}Fractal dimension (FD) of different eigenstates as
a function of corresponding energies $E$ and perturbative on-site potential $v_{p}$. with $m=1$, $\lambda=2.0$, $t_0=1.0$ and $L=2584$.}
\end{figure}

\section{More perturbation results}\label{MorePerturbation}
In the main text, we have showed that the critical zone is robust against experimentally relevant onsite perturbation and is also against interactions between particles. To further demonstrate robust critical zone is indeed protected by zeros of hopping coefficients, we present more numerical results of perturbations. In this section, the on-site perturbation $V_{p}$ is given by
\begin{equation}
V_{p}=v_{p}\sum_{j}\cos(2\pi m\alpha j+\phi)n_{j},
\end{equation}
with $m\in \mathbb{Z}$, and the hopping perturbation $T_{P}$ is given by
\begin{equation}
T_{p}=t_{p}\sum_{j}[\cos(2\pi m\alpha j+\phi)b_{j}^{\dagger}b_{j+1}+\mathrm{H.c.}],
\end{equation}

\subsection{Simplest onsite perturbation}
We first consider the onsite perturbation $V_{p}$ with $m=1$. Fig.~\ref{fig:FDSMonsite} shows FD of eigenstates as a function of energies
$E$ and strength of perturbations $v_{p}$. The critical zone persists for a range of perturbation $v_{p}$. As $v_{p}$ increases, the critical zone shrinks since the on-site perturbation localizes some of the critical states. When $v_{p}$ is large enough, all the critical states will be localized as expected. This result is consistent with the experiment-relevant perturbation in the main text.

\subsection{More perturbations}
\begin{figure}[thp]
\includegraphics{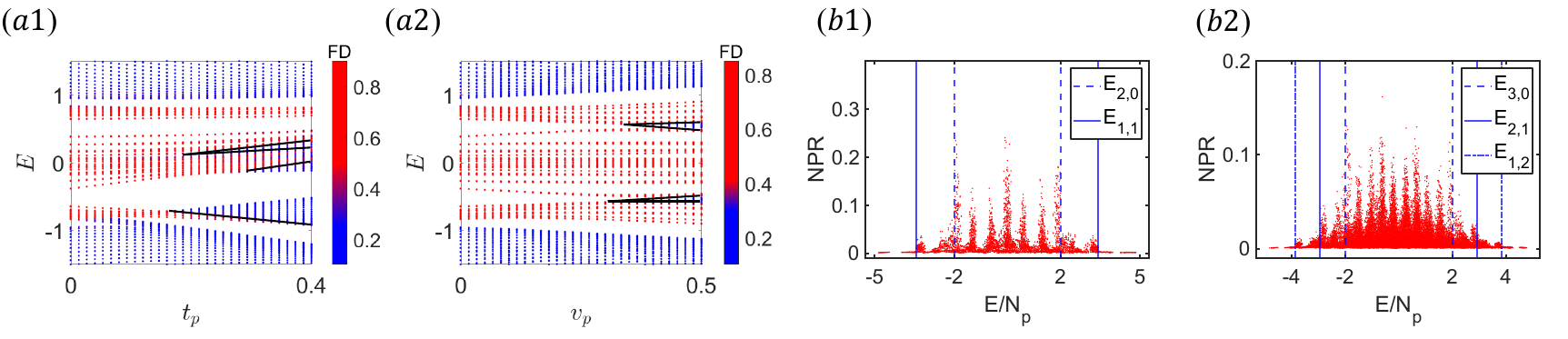}\caption{\label{fig:FDSM}Fractal dimension (FD) of different eigenstates as
a function of corresponding energies $E$ and (a1) strength of perturbative
hopping term $t_{p}$ and (a2) perturbative on-site potential $v_{p}$. New MEs are marked by black lines. Other parameters are $\lambda=0.8$ and $L=2584$. And Normalized participation ratio (NPR) as a function
of energy density $E/N_{\mathrm{p}}$ of few hardcore bosons case with nearest neighbor interaction $U=1.4$, with $\lambda=2.0$, $t_0=1.0$ and (b1) $N_{\mathrm{p}}=2$, $L=120$ (b2) $N_{\mathrm{p}}=3$, $L=60$. The critical energy $E_{2,0}$($E_{3,0}$) is equal to the single particle ME $E=\lambda$ for two (three) bosons case.}
\end{figure}

Moreover, we choose $m=3$ and consider both onsite perturbation and perturbative hopping term. Fig.~\ref{fig:FDSM} shows FD of eigenstates as a function of energies
$E$ and strength of perturbations and interactions. Similar to the results in the
main text and last subsection, the critical zone is robust against moderate perturbation and can persist for a range of $t_p$ or $v_p$ as expected. Furthermore, it is observed that adding perturbative hopping term $T_p$ is easier to drive system into localized
phase compared to adding perturbative on-site term $V_p$. This is because $T_p$ directly acts on the hopping terms and is easier to remove the zeros of hopping coefficients compared to the on-site perturbation $V_p$. Another feature is that a second ME emerges due to perturbations, and the localization transitions start from center of the spectrum instead of edge of the spectrum.
One can introduce more terms as perturbation and choose $m$ as arbitrary
integer and will obtain similar results, indicating robustness of
the MEs protected by the generic mechanism.

We also present more numerical results of robustness against interactions. For the case $\lambda=2$, $t_0=1$ and $V=1.4$, the critical energy is less prominent compared with the results in the main text, but the relation $E_{c,l}=(c\lambda+l E_\mathrm{max})$ still holds and single particle ME still exists as illustrated in Fig.~\ref{fig:FDSM}(b).

\section{Localization length}\label{LocLen}
\begin{figure}[thp]
\includegraphics{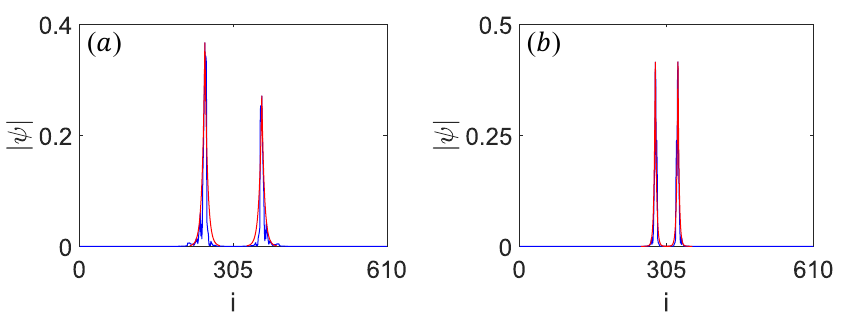}\caption{\label{fig:lengthSM}Spatial distributions of eigenstates. Blue lines represent $|\psi|$, where $\psi$ is the eigenstates corresponding to (a) $E=-2.1(6)t_0$ with $\lambda=2t_0$, (b)$E=-1.3(5)t_0$ with $\lambda=t_0$. Here we fix $L=610$. The red lines represent $|\psi|_{\mathrm{max}}\exp{(-|i-i_0|/\xi)}$.}
\end{figure}

In this section, we numerically verify the localization length $\xi$ obtained in the main text. As shown in Fig.~\ref{fig:lengthSM}, the red lines represent $|\psi|_{\mathrm{max}}\exp{(-|i-i_0|/\xi)}$, where $|\psi|_{\mathrm{max}}$ is the maximum values of $|\psi|$ in the two peaks, $i_0$ is the corresponding lattice sites and $\xi$ is the localization length which is given by $\xi(E)=2/\ln[(|E|+\sqrt{E^{2}-\lambda^{2}})/|\lambda|]$. It is indicated the analytic experssions of the localization length $\xi$ well describe the localization features of the corresponding eigenstates.

\section{Experimental Realization}\label{ExpReal}
In this section, we illustrate how to realize the lattice model in the main text. Our basic idea is to use well-tuned Rydberg atoms to simulate hard-core bosons with hoppings and on-site potentials controlled by external fields. We shall realize the Hamiltonian
\begin{equation}
H=  \sum_{j_x}(J_{j_x}\sigma_{j_{x},A}^{+}\sigma_{j_{x},B}^{-}+\lambda \sigma_{j_{x},A}^{+}\sigma_{j_{x}+1,B}^{-}+\mathrm{H.c.})+\frac{1}{2}\sum_{j_{x},s=\{A,B\}}V_{j_x}\left(\mathbb{1}+\sigma_{j_{x},s}^{z}\right),
\end{equation}
which can be transformed into the model in the main text by relabeling site index $j_{x}\rightarrow 2j-1$ for $j_x$ on A-leg and $j_{x}\rightarrow 2j$ for $j_x$ on B-leg, and defining bosonic operator $b_{j}^{\dagger}=\lvert 1\rangle_{j}\langle 0\rvert_{j}$
at each site, with $\lvert 1\rangle\equiv\lvert \uparrow\rangle$ and $\lvert 0\rangle\equiv\lvert \downarrow\rangle$. In the following we shall take $^{87}$Rb atoms as an example. For $^{87}$Rb, two Rydberg states are chosen to simulate spin $1/2$ at each site with $\lvert\downarrow\rangle\equiv\lvert 70,S\rangle$
and $\lvert\uparrow\rangle=\lvert 70,P\rangle$. We consider the two-leg superarray of Rydberg atoms,
with each trapped in optical tweezers. We consider the two-leg superarray of Rydberg atoms,
with each trapped in optical tweezers. An effective Zeeman splitting $M_{{j_x},s}$ is introduced along the $x$ direction, with $M_{{j_{x}+1},s}-M_{{j_{x}},s}=\Delta$, $M_{{j_{x}=0},A}=0$ and $M_{{j_{x}=0},B}=0$. Three Raman beams are applied to generate Raman and AC Stark potential. The total Hamiltonian of the system is given by,
\begin{equation}
H=H_{\mathrm{dipole}}+H_{\mathrm{Zeeman}}+V_{R}(\boldsymbol{r},t)+V_{\mathrm{AC}}(\boldsymbol{r}),
\end{equation}
which includes the bare dipole-dipole interactions~\cite{rydberg1sm}
\begin{equation*}
H_{\mathrm{dipole}}=\sum_{j_{x}}J_{y}^{0}\sigma_{j_{x},A}^{+}\sigma_{j_{x},B}^{-}+J_{d}^{0}\sigma_{j_{x}-1,A}^{+}\sigma_{j_{x},B}^{-}+\sum_{|i-j|>1}\frac{J_{ij}}{R_{ij}^{3}}\sigma_{i,A}^{+}\sigma_{j,B}^{-}+\mathrm{H.c.},
\end{equation*}
with $J_{y}^{0}=C_3/a_{y}^{3}$ and $J_{d}^{0}=C_{3}/(a_{x}^{3}+a_{y}^{3})^{3/2}$, the effective Zeeman energy gradient term,
\begin{equation*}
H_{\mathrm{Zeeman}}=\frac{1}{2}\sum_{j_{x},s=\{A,B\}}{M_{j_x,s}}\sigma_{j_{x},s}^{z},
\end{equation*}
the Raman coupling potential obtain from time-dependent perturbation~\cite{laddism}
\begin{equation*}
V_{R}(\boldsymbol{r},t)=\frac{\Omega_{1}(\boldsymbol{r},t)^{*}\Omega_{2}(\boldsymbol{r},t)}{\Delta_{i}}\sigma^{x}_{j_{x},j_{y}},
\end{equation*}
and the AC Stark term
\begin{equation*}
    V_{\mathrm{AC}}(\boldsymbol{r})=\frac{|\Omega_{3}(\boldsymbol{r})|^{2}}{\Delta_{i}'}.
\end{equation*}
As discussed in the main text, the bare dipole-dipole interactions between Rydberg states with the same Zeeman energy constitute the constant hopping coupling $\lambda=J_{d}^0$. In the following, we discuss the scheme for realizing the incommensurate hopping couplings and on-site potential.
\subsection{The incommensurate hoppings}
The incommensurate hoppings are simulated by laser-assisted dipole-dipole interactions. The bare dipole-dipole interactions between legs on the same $j_x$ are
suppressed by the large Zeeman splitting $\Delta$, but can be further
restored by applying the Raman coupling potential $V_{R}$. The Raman
coupling potential is generated by two Raman lights with Rabi-frequencies
$\Omega_{1,2}$ and the frequencies $\omega_{1,2}$. When the frequency
difference satisfies nearly resonant condition $\omega_{1}-\omega_{2}\approx\Delta$,
then the Zeeman detuning $\Delta$ is compensated by the two-photon
process. Here we choose $\Omega_{1}=\Omega_{0}\cos(2\pi\alpha j_{x}+\frac{\pi}{4})e^{i\theta_{1}}$
and $\Omega_{2}=\Omega_{0}\sin(2\pi\alpha j_{x}+\pi j_{y}+\frac{\pi}{4})e^{i\theta_{2}}$,
direct calculation yields the Raman potential
\begin{equation}
  V_{R}(\boldsymbol{r},t)=\sum_{j_{x},j_{y}}\frac{|\text{\ensuremath{\Omega_{0}}}|^{2}}{2\Delta_{i}}\cos(4\pi\alpha j_{x})\cos(\pi j_{y})e^{i\delta\theta}\sigma_{j_{x},j_{y}}^{x}+\mathrm{H.c.},
\end{equation}
with $\delta\theta=\theta_{2}-\theta_{1}$. So we choose $\theta_{2}=\theta_{1}+n\pi$,
which can be realized in experiment. To the lowest order, the effective exchange couplings are
~\cite{laddism}
\begin{equation}
J_{j_{x}}=\frac{|\text{\ensuremath{\Omega_{0}}}|^{2}}{\Delta\Delta_{i}}\cos(2\pi\alpha j_{x})J_{y}^{0},
\end{equation}
which realize the incommensurate hopping coefficients. Notice that the transition between $j_{x}$ and $j_{x}+2$ and longer
distance are suppressed by larger Zeeman energy offset (i.e., at least
$2\Delta$) and cannot be induced by two-photon process.
To prohibit laser-assisted exchange coupling within the same leg, we use the angular dependence of dipole-dipole
interaction $J=d^{2}(3\cos^{2}\theta_{ij}-1)/R_{ij}^{2}$ as discussed in the main text. We arrange the Rydberg atoms along the "magic angle" $\theta_{m}=\arccos(1/\sqrt{3})\approx54.7^{\circ}$, at which the dipolar interactions vanish. So there is no hopping coupling generated by Raman potential along the $x$ direction.

\subsection{The incommensurate on-site potential}
We apply another standing wave field to generate an incommensurate on-site potential $V_\mathrm{AC}(\boldsymbol{r})$, which is $\Omega_3(\boldsymbol{r})=\Omega_{0}'\cos(2\pi\alpha j_{x})$. Together with an additionally correction from Raman potential, the total on-site term is given by
\begin{eqnarray}
V_{\mathrm{on-site}}&=&\frac{|\Omega_{1}|^{2}+|\Omega_{2}|^{2}}{\Delta_{i}} + V_{\mathrm{AC}}\nonumber\\
&=&\frac{1}{\Delta_{i}}\left[|\Omega_{0}|^{2}\cos^{2}(2\pi\alpha j_{x}+\frac{\pi}{4})+|\Omega_{0}|^{2}\sin^{2}(2\pi\alpha j_{x}+\pi j_{y}+\frac{\pi}{4})\right]+\frac{1}{\Delta_{i}'}\left[|\Omega_{0}'|^{2}\cos^{2}(2\pi\alpha j_{x})\right]\nonumber\\
&=&\frac{|\Omega_{0}|^{2}}{\Delta_{i}} + \frac{|\Omega_{0}'|^{2}}{2\Delta_{i}'} + \frac{|\Omega_{0}'|^{2}}{2\Delta_{i}'} \cos(4\pi\alpha j_x),
\end{eqnarray}
of which the constant part can be neglected and we can reach the desired on-site incommensurate potential.

\subsection{Experimental parameters}
Here, we also give an estimate of the orders of magnitude of the relevant experimental parameters. The data given here are based on $^{87}$Rb atoms. For the states $\left|70S\right>$ and $\left|70P\right>$, $C_3 = 4196 \text{MHz} \cdot \mu m^{-3}$. Then with $a_y/a_x=0.8$, $\left|\Omega_0\right| = \left|\Omega_0'\right| = 200$ MHz, $\Delta = 10$MHz and $\Delta_i = 16.5$Ghz and $a_x = 12.5 \mu m$, we obtain the model with $t_0 = \left|\Omega_1\Omega_2 J^{0}_{y}/(2\Delta \Delta_i)\right| = 0.5$ MHz, $\lambda = 1$MHz

To observe the correlated effects, the lifetime of the system should be large compared to the characteristic time of the system. For $\lambda = 1$MHz, a lifetime of at least $\tau > \tau_{0} = 100J_0^{-1} = 100\mu s$ is needed. The lifetime of the Rydberg states $\left|70S\right>$ and $\left|70P\right>$ are $370\mu s$ and $750\mu s$ respectively. The lifetime of $6P$ states, which the Raman process coupled, has a lifetime of $\tau_{6P}=130 ns$. So for the single photon detuning $\Delta_i = 16.5$ GHz and $\Delta_i' = 20$ GHz, the lifetime can be estimated by $\tau = \left(2|\Omega_0|/\Delta_i + |\Omega_0'|/\Delta_i'\right)^{-2} \tau_{6P} \approx 110 \mu s > \tau_0$.

\end{document}